\documentclass[fleqn,usenatbib]{mnras}

\usepackage{newtxtext,newtxmath}

\usepackage[T1]{fontenc}

\DeclareRobustCommand{\VAN}[3]{#2}
\let\VANthebibliography\thebibliography
\def\thebibliography{\DeclareRobustCommand{\VAN}[3]{##3}\VANthebibliography}

\newcommand{\kms}{km\,s$^{-1}$} 
\newcommand{\HI}{H{\sc i}}

\usepackage{graphicx}	
\usepackage{amsmath}	

\makeatletter

\newcommand{\Rmnum}[1]{\expandafter\@slowromancap\romannumeral #1@}
\makeatother

\title[The obstructed jet in Mrk~231]{The obstructed jet in Mrk~231}

\author[Wang et al.]{
Ailing Wang,$^{1,2}$
Tao An,$^{1}$\thanks{E-mail: antao@shao.ac.cn}
Sumit Jaiswal$^{1}$, 
Prashanth Mohan$^{1}$, 
Yuchan Wang$^1$,
\newauthor
Willem A. Baan$^{3,1}$, 
Yingkang Zhang$^1$,
Xiaolong Yang$^1$
\\
$^{1}$Shanghai Astronomical Observatory, Key Laboratory of Radio Astronomy, CAS, 80 Nandan Road, Shanghai 200030, China \\
$^{2}$University of Chinese Academy of Sciences, 19A Yuquanlu, Beijing 100049, China \\
$^{3}$Xinjiang Astronomical Observatory,  Key Laboratory of Radio Astronomy, CAS, 150 Science 1-Street, Urumqi, Xinjiang 830011, China \\
}

\date{Accepted XXX. Received YYY; in original form ZZZ}

\pubyear{2015}

\begin{document}
\label{firstpage}
\pagerange{\pageref{firstpage}--\pageref{lastpage}}
\maketitle

\begin{abstract}
Mrk~231 is the closest radio-quiet quasar known and one of the most luminous infrared galaxies in the local Universe. It is characterised by the co-existence of a radio jet and powerful multi-phase multi-scale outflows, making it an ideal laboratory to study active galactic nucleus (AGN) feedback. We analyse the multi-epoch very long baseline interferometry data of Mrk~231 and estimate the jet head advance speed to be $\lesssim0.013\ c$, suggesting a sub-relativistic jet flow. The jet position angle changes from $-113\degr$ in the inner parsec to $-172\degr$ at a projected distance of 25 parsec. The jet structure change might result from either a jet bending following the rotation of the circum-nuclear disc or the projection of a helical jet on the plane of the sky. In the large opening angle ($\sim60\degr$) cone, the curved jet interacts with the interstellar medium and creates wide-aperture-angle shocks which subsequently dissipate a large portion of the jet power through radiation and contribute to powering the large-scale outflows. The low power and bent structure of the Mrk~231 jet, as well as extensive radiation dissipation, are consistent with the obstruction of the short-length jet by the host galaxy's environment.
\end{abstract}

\begin{keywords}
galaxies: ISM -- galaxies: jets -- galaxies: kinematics and dynamics -- quasars:
individual (Mrk 231)
\end{keywords}



\section{Introduction}

Mrk~231, also known as UGC 08058 and IRAS 12540+5708 at a redshift $z =  0.042$ \citep{1991rc3..book.....D}, is a relatively nearby  ultra-luminous infrared galaxy (ULIRG) \citep{1988ApJ...325...74S}.  Discovery of strong hydroxyl (OH) megamasers in this galaxy supports an extreme starburst in the nuclear region \citep{1985Natur.315...26B,2003Natur.421..821K}, consistent with its high infrared luminosity of $\approx 3 \times 10^{12}~ L_\odot$ \citep{1989AJ.....98..766S}. 

Mrk~231 is characterised by prominent multi-phase outflows extending over multiple physical scales \citep{2014ApJ...788..123L}. Optical broad absorption lines (BALs) are detected in both low and high ionisation species and indicate galactic-scale outflows arising at the pc-scales \citep{2013ApJ...764...15V,2016ApJ...825...42V}. The study of \citet{2015A&A...583A..99F} identifies a molecular gas outflow with a velocity $\geq 400$ \kms originating from the nuclear region and extending to $\approx 1$ kpc, and an ionised X-ray ultra-fast outflow (UFO) with a velocity of $\approx$ 20000 \kms originating from the accretion disc; an analysis (based on their energetics and momentum transfer) suggests that a bulk of the UFO kinetic energy is transferred to the mechanical energy of the kpc-scale molecular outflow. The study of \cite{2011ApJ...729L..27R} reports on the detection of fast ($\approx$ 1000 \kms) outflowing neutral gas (as traced by the Na I D absorption) extending from the nuclear region up to 3 kpc; they suggest that this is driven by radiative pressure or mechanical energy from the central AGN. Neutral and ionised gas outflows have also been observed with velocities up to $10^3$ \kms\  \citep{2005ApJ...632..751R,2011ApJ...729L..27R,2017ApJ...850...40R}. \HI\ 21 cm absorption features are detected \citep{1998AJ....115..928C, 2016A&A...593A..30M} on scales ranging from a few to a few hundred pc. The \HI\ line emission traces an east-west elongated disc possibly coincident with the inner part of the molecular disc at an even larger scale. 

The multi-phase multi-scale outflows in active galactic nuclei (AGN) may be enabled by a mixture of driving mechanisms, including the central AGN (radiation pressure), a nuclear starburst (stellar winds and radiation pressure), and the radio jet (kinetic energy). While the role of AGN-driven winds and outflows (non-relativistic) on feedback at large scales has been investigated in some detail \citep[e.g., ][]{2017ApJ...850...40R,2021A&A...645A..21G}, the role of the radio jet in driving the outflow in the nuclear region requires clarification \cite[e.g.][]{2013Sci...341.1082M}. Mrk~231 is one of the few radio-quiet quasars containing a prominent jet. The source shows a triple (core and two lobes) radio morphology structure with an extent of $\sim$40 pc (50 mas) in the north-south direction, revealed by the European VLBI Network (EVN) observation at 1.6 GHz \citep{1988AJ.....96..841N} and the Very Long Baseline Array (VLBA) observations at multiple frequencies from 1.4 to 8.4 GHz \citep{1999ApJ...516..127U}. The higher-resolution VLBA images at 15 GHz resolve the central component into two compact components separated by $\sim$1 pc aligning in the northeast-southwest direction \citep{1999ApJ...516..127U,1999ApJ...517L..81U}. The northeastern component showed a remarkable variability \citep{1999ApJ...517L..81U} and was thus identified as the core (the location of the active nucleus). The inner 1-pc position angle of the jet is different from that of the larger 40-pc triple structure by $\sim60\degr$. \citet{2009ApJ...706..851R,2017ApJ...836..155R,2020ApJ...891...59R} observed Mrk~231 using the VLBA at higher frequencies of 22 and 43 GHz and obtained a similar core-jet morphology at 1-pc scale. The derived core brightness temperatures, $T_{\rm B} > 10^{10}$ K, are 2--3 orders of magnitude lower than that of typical blazars \citep{2005AJ....130.2473K}, but are consistent with radio-quiet quasars \citep{1996ApJ...468L..91B,2005ApJ...621..123U}. The kinematic property of the VLBI jet is still under debate. \citet{1999ApJ...517L..81U} determined a jet proper motion speed of $0.14 \pm 0.052\, c$ in the period from 1996.8 to 1998.7, while \citet{2017ApJ...836..155R} measured a higher jet apparent speed $>3.15 \,c$ during a flaring state in 2015. This can be explained if Mrk~231 contains episodic fast-moving ($v \approx 1-3\, c$) knots and a continuous slower background jet body \citep{2020ApJ...891...59R}. 

In this paper, we use archival VLBI data of Mrk~231 to study the jet kinematics. We discuss the complex nature of the pc-scale jet and possible connection between the wide-angle outflows and the radio jet. Throughout, we adopt the following cosmological parameters: H$_0$ = 71 \kms\ Mpc$^{-1}$, $\Omega_\Lambda = 0.73$ and $\Omega_m = 0.27$. At $z = 0.042$, 1 mas angular separation corresponds to 0.818 pc linear size in projection on the plane of the sky, and 1 mas yr$^{-1}$ jet proper motion is converted to $2.78\,c$.

\section{Observations and data reduction}

\begin{table*}
    \centering
    \caption{Logs of the 8.4 GHz VLBI observations.}
    \begin{tabular}{cccccccc}
    \hline \hline 
    Date         & Project Code & Bandwidth  & On-source time &Beam                             & $\sigma_{\rm rms}$ & $S_{\rm peak}$   \\
    (yyyy-mm-dd) &              & (MHz)      &(min)           & (maj, min, PA) & (mJy beam$^{-1}$)  &(mJy beam$^{-1}$) \\ 
    \hline 
    1994-08-12   &BB023         &16          &3.9             &$1.9\times0.9$ mas$^2$, $15.9\degr$       &1.02                &82  \\
    1998-09-15   &BU013         &32          &37.7            &$1.8\times1.0$ mas$^2$, $14.8\degr$       &0.14                &117 \\ 
    1999-03-08   &RDV13         &32          &25.2            &$1.0\times0.8$ mas$^2$, $-47.5\degr$      &0.26                &86  \\
    1999-12-16   &BU013         &16          &11.2            &$3.0\times1.2$ mas$^2$, $32.2\degr$       &0.16                &135 \\
    2000-01-08   &BU013         &16          &13.3            &$3.0\times1.1$ mas$^2$, $57.4\degr$       &0.21                &138 \\
    2006-09-03   &BA080         &32          &3.2             &$3.5\times1.0$ mas$^2$, $49.5\degr$       &0.36                &156 \\
    2014-06-09   &BG219D        &384         &4.0             &$1.5\times1.2$ mas$^2$, $-77.0\degr$      &0.16                &120 \\
    2017-01-21   &UF001B        &384         &4.7             &$1.1\times1.1$ mas$^2$, $4.0\degr$        &0.20                &142 \\
    2018-12-04   &UG002U        &384         &6.4             &$1.4\times1.1$ mas$^2$, $15.7\degr$       &0.15                &288 \\
    \hline
    \end{tabular} \\
    Note: Column 5 gives the restoring beam; Columns 6 and 7 present the {\it rms} noise and peak intensity in the image, respectively. 
    \label{table1}
\end{table*}

\begin{table*}
    \centering
    \caption{Model fitting results.}
    \begin{tabular}{ccccccc}
    \hline \hline 
    Time         & Comp. & $S_{\rm int}$ & R                  & PA               &$\theta_{\rm FWHM}$ &$T_{\rm B}$   \\
    (yyyy-mm-dd) &       & (mJy)         &(mas)               &(deg)             &(mas)             &(10$^{10}$ K)   \\ 
    (1) & (2) & (3) & (4) & (5) & (6) & (7) 
    \\ \hline 
    1994-08-12   &J1      &92$\pm$14      &1.086$\pm$0.023       &$-112.4\pm0.9$    &0.43$\pm$0.05     & 0.9$\pm$0.3   \\
                 &C       &48$\pm$7       &0                   &0                 &0.21$\pm$0.03     & 2.0$\pm$0.4       \\
    1998-09-15   &J1      &132$\pm$9      &1.154$\pm$0.014       & $-111.7\pm0.7$   &0.45$\pm$0.03     & 1.2$\pm$0.2   \\
                 &C       &27$\pm$2       &0                   &0                 &0.10$\pm$0.02     & $>4.9$        \\
    1999-03-08   &J1      &96$\pm$8       &1.117$\pm$0.015       & $-112.4\pm0.8$   &0.29$\pm$0.02     & 1.9$\pm$0.3   \\
                 &C       &49$\pm$5       &0                   &0                 &0.10$\pm$0.02     & $>8.3$        \\
    1999-12-16   &J1      &130$\pm$9      &1.032$\pm$0.015       & $-115.6\pm0.8$   &0.41$\pm$0.03     & 1.4$\pm$0.2   \\
                 &C       &28$\pm$2       &0                   &0                 &0.10$\pm$0.02     & $>5.0$        \\
    2000-01-08   &J1      &121$\pm$9      &1.024$\pm$0.014       & $-113.6\pm0.8$   &0.36$\pm$0.02     & 1.6$\pm$0.2   \\
                 &C       &33$\pm$2       &0                   &0                 &0.10$\pm$0.02     & $>5.9$        \\
    2006-09-03   &J1      &149$\pm$12     &1.154$\pm$0.014       &$-113.5\pm0.7$    &0.51$\pm$0.04     & 1.1$\pm$0.2   \\
                 &C       &51$\pm$4       &0                   &0                 &0.10$\pm$0.02     & $>6.6$        \\
    2014-06-09   &J2      &124$\pm$9      &0.978$\pm$0.014       &$-116.5\pm0.8$    &0.47$\pm$0.03     & 1.0$\pm$0.2   \\
                 &C       &36$\pm$3       &0                   &0                 &0.10$\pm$0.02     & $>6.1$        \\
    2017-01-21   &J2      &155$\pm$12     &1.006$\pm$0.015       &$-118.3\pm0.8$    &0.39$\pm$0.03     & 1.7$\pm$0.3   \\
                 &C       &98$\pm$8       &0                   &0                 &0.10$\pm$0.02     & $>16.5$       \\
    2018-12-04   &J2      &291$\pm$17     &0.982$\pm$0.013       &$-115.9\pm0.8$    &0.36$\pm$0.02     & 3.8$\pm$0.5   \\
                 &C       &117$\pm$7      &0                   &0                 &0.10$\pm$0.02     & $>19.9$       \\
    \hline
   \end{tabular}\\
Columns (3) to (7) present in sequence the integrated flux density, radial separation, and position angle with respect to the core C, component size (full width at half maximum of the fitted Gaussian component) and brightness temperature. 
   \label{table2}
\end{table*}

We obtained the 8-GHz VLBA data at nine epochs from the public archive of the National Radio Astronomy Observatory (NRAO) of the US\footnote{\url{https://science.nrao.edu/observing/data-archive}} and  Astrogeo Center\footnote{\url{http://astrogeo.org/} maintained by L. Petrov.}. The data covers a 25-yr long interval from 1994 to 2019. The observational setup is presented in Table~\ref{table1}. The NRAO VLBA data was calibrated and analysed in Astronomical Image Processing System \citep[AIPS,][]{Greisen2003} following the standard procedure described in the cookbook. The Astrogeo data has been calibrated using the VLBI processing software PIMA \citep{Petrov_2011}. The calibrated visibility data sets were further analysed in the Caltech DIFMAP software package \citep{1995BAAS...27..903S} to handle the residual phase errors and imaging. Data with significant scatter were averaged over a time span of 30 s. Bad data points appearing as outliers were mostly caused by the radio frequency interference and thus flagged. The edited data was then self-calibrated and mapped following an automated pipeline and cross-verified through manual interaction. Natural weighting was utilised to create the final maps to obtain higher sensitivity. The images were de-convolved with the restoring beam using the CLEAN algorithm \citep{1974A&AS...15..417H}. The CLEAN-ing was done up to three times the rms noise  level of the residual map.

We used circular Gaussian models to fit the self-calibrated visibility data and estimated the integrated flux densities and angular size of the core and jet components. First, a single Gaussian component was fitted on the source. Next, we extracted the model component from the data and created the residual map, from which we continued to fit a second component with a circular Gaussian model. Then, we took these two circular Gaussian components together to fit the original visibility data to ensure that the model best matches the data. The model fitting stopped when there was no visible feature brighter than 5$\sigma$ appearing in the residual map. The archive data was mostly from astrometric or geodetic observations carried out in snapshot mode. The on-source time ranged from 3.2 to 37.7 minutes, and the acquired images were of varying qualities (see Table \ref{table1}). Therefore, we only focused on these two bright components, and did not attempt to fit other weaker emission structures. We also used two elliptical Gaussian models to fit the visibility data and found consistency with the circular Gaussian models. Based on previous VLBI observations \citep{1999ApJ...517L..81U,2009ApJ...706..851R,2017ApJ...836..155R,2020ApJ...891...59R}, the northeast component is identified as the core, and the southwest component is a jet knot. The model-fitting results, including the flux density, size, radial separation with respect to the core and brightness temperature are presented in Table \ref{table2}. 

The uncertainties in the model fitting parameters were estimated using the relations given in \citet{2012A&A...537A..70S}. The calculations were based on the approximations introduced by \citet{1999ASPC..180..301F} and were modified for the strong sidelobes induced by incomplete VLBI ($u,v$) coverage. This modified error estimate \citep{2012A&A...537A..70S} was suitable for the current data obtained from snapshot observations and thus had large sidelobes. When the fitted Gaussian size $d$ of a component was smaller than the minimum resolvable size $d_{\rm min}$ of a given interferometer, $d_{\rm min}$ was used in place of $d$. $d_{\rm min}$ can be estimated using the equation given by \citet{2005astro.ph..3225L}:
\begin{equation}
    d_{\rm min}=\cfrac{2^{2-\beta}}{\pi}\left[\pi a b \ln2\ln\left(\cfrac{\rm SNR}{\rm SNR-1}\right)\right]^{1/2}
\end{equation}
where $a$, $b$ are the major and minor axes sizes of the point spread function (the synthesised beam), SNR is the signal-to-noise ratio of the component's peak brightness to the image noise, $\beta$ is the power-law index describing the visibility data weighting, i.e., $\beta = 0$ is used for the uniform weighting and $\beta = 2$ for the natural weighting. The uncertainties in the component sizes were estimated as the square-root of quadratic sum of their corresponding statistical and fitting errors. The statistical error in component size was estimated using the relation given in \citet{2012A&A...537A..70S}. The fitting error in component size, on the other hand, was estimated by varying the flux density and size parameters in small steps around the initial fitted values so that the fitted parameters in each step of the modified model were not very different from the initial values and the corresponding reduced chi-squares were close to one. During the fitting all parameters but the model type were kept variable. The standard deviation of the fitted component sizes obtained after each fitting iteration of the most offset modified model was taken as the fitting error in component size. The position errors were then estimated as the half of the component size errors. 

\section{Results}\label{sec:3}

\subsection{Radio morphology}\label{sec:3.1}
\begin{figure}
    \centering
    \includegraphics[width=0.5\textwidth]{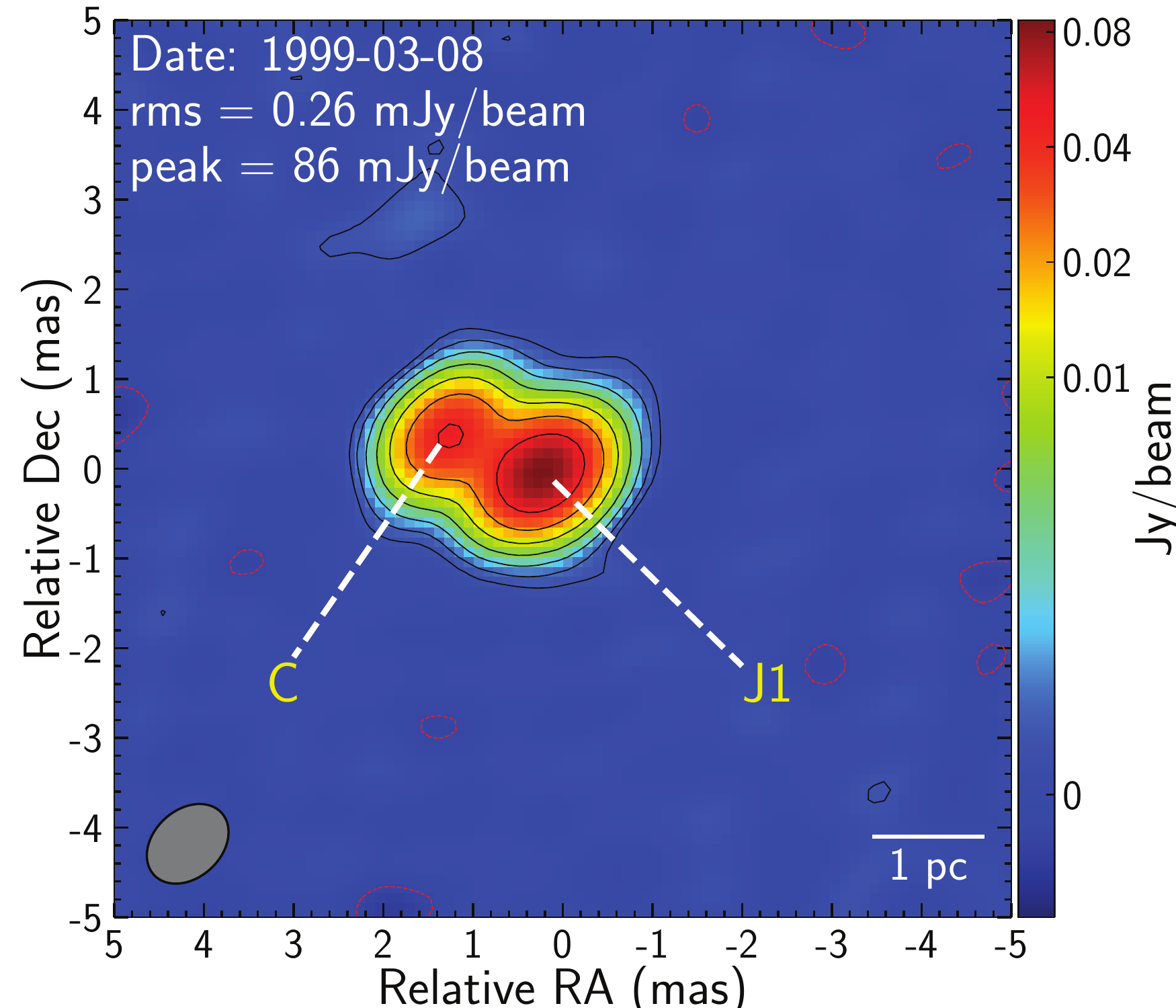}
    \includegraphics[width=0.5\textwidth]{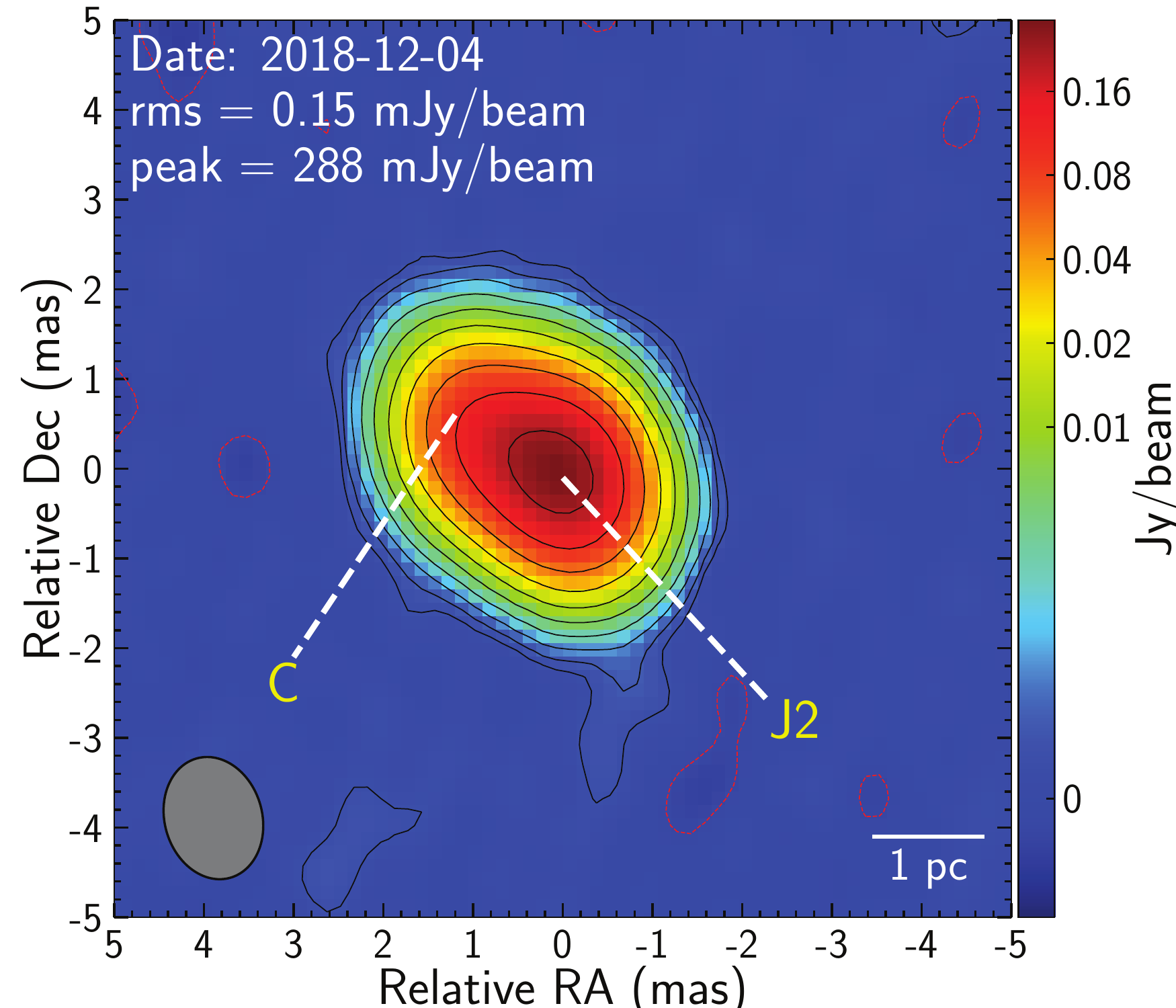}
    \caption{VLBI images of Mrk~231 observed at 8~GHz on the epoch 1999 March 8 and 2018 December 4. The images are created with natural weighting. The negative red coloured contour is at $-2\sigma$ level and the positive black-coloured contour levels are in the series of 3$\sigma \times$(1, 2, 4, 8, 16, 32, 64), where $\sigma$ is the \textit{rms} noise in each image. The colour scale shows the intensity in the logarithmic scale.}
    \label{VLBI}
\end{figure}

Figure \ref{VLBI} shows the 8-GHz VLBI images of Mrk~231 observed on 1999 March 8 and 2018 December 4. The images of other epochs show a similar morphology but a lower quality and thus are not shown here. The image parameters are presented in Table \ref{table1}. The emission structure has a largest extent of about two mas (corresponding to a projected size of about 1.6 pc) along the northeast--southwest direction, consistent with the previous VLBI observational results \citep[e.g., ][]{1999ApJ...517L..81U,2017ApJ...836..155R}. The northeastern component was identified as the core in previous studies \citep{1999ApJ...517L..81U,2009ApJ...706..851R}. We use the same nomenclature in this paper. In the 8-GHz images, the core is weaker but more compact than the jet component (see Table \ref{table2}); however, the core is brighter at higher frequencies ($\geq 15$ GHz), especially during flaring periods \citep[e.g., ][]{2017ApJ...836..155R}). 

The jet component's peak emission in the 8 GHz images is about 1.2 mas ($\sim$1 pc) away from the core. The jet position angle (PA) and the core--jet separation (R) are plotted as a function of time in Figure~\ref{fig2}. We find that the jet features are not easily described with a single component from the image. Instead, two groups of jet components may be inferred: the components between 1994 and 2007 have similar separation ($\sim1.15$ mas) and position angle ($\sim-113\degr$), and are labelled as J1; other components between 2014 and 2019 are labelled as J2 (see Table \ref{table2}). \citet{2017ApJ...836..155R,2020ApJ...891...59R} also detected a jet component K1 at a similar distance and position angle with J2 during the 2015 and 2017 flares. The authors assumed K1 is a stationary component and determined the relative motion of the core toward K1, $0.97\, c$. This motion is  attributed to a newly discrete jet ejection that also dominates the total nuclear flux density at 43 GHz. Determining this small displacement ($\sim$0.075 mas) in the inner jet is only possible by virtue of the high-accuracy astrometry achieved by their phase-referencing observations and the high-resolution VLBI images at 43 GHz. In contrast, J1 and J2 in our study are detected from lower-frequency, lower resolution 8-GHz images. Moreover, the 8.4-GHz core suffers from synchrotron self-absorption that hinders the detection of minor changes in the emission structure of the core region. Therefore, our present study focuses on the kinematics of the underlying jet.

\citet{1999ApJ...516..127U} reveals a north-south elongated triple structure with a total extent of 40 pc in their 1.4-GHz VLBA image. The southern lobe is brighter and longer, thus is identified to be associated with the advancing jet. The peak of the southern lobe is $\sim$25 pc ($\sim30$ mas) away from the core at a position angle of $\sim -172\degr$. Figure~\ref{VLBI} displays the inner-parsec jet extending to the southwest direction at a position angle $\sim -115\degr$. The misalignment between the inner 1-pc and the outer 25-pc southern jet is about $60\degr$, and the mean position angle of the VLBI jets is $-144\degr$. The jet geometry is consistent with that of the molecular outflow. The molecular CO (2--1) outflow shows a wide-angle biconical geometry with a size of $\sim$1 kpc along the northeast (red-shifted component) to southwest (blue-shifted component) directions \citep{2015A&A...583A..99F}. In the nuclear region, the OH megamaser emission reveals a dusty molecular torus or thick disc located between 30 and 100 pc from the central engine \citep{2003Natur.421..821K}. The dusty disc's projected position angle is $-145\degr$, in a good alignment with the central axis of the cone cleaned by the jet. The radio jet seems to follow a twisted path within the conical cone and is prominent at the boundaries of the cone. 

Both the core and jet components show a compact morphology. The brightness temperatures ($\mathrm{T}_{\rm B}$) of the VLBI components are calculated using the equation 2 of \citet{1982ApJ...252..102C}. The core brightness temperature is remarkably higher than the jet brightness temperature (Table~\ref{table2}), reinforcing the core identification. We should note that, except for the first epoch, the core in the other epochs is unresolved; the size ($\theta_{\rm FWHM}$) is given as an upper limit, and thus the $\mathrm{T}_{\rm B}$ represents a lower limit. The core brightness temperature of Mrk~231 exceeds $10^{10}$~K, substantially higher than typical values for radio-quiet AGN \citep{2005ApJ...621..123U,1996ApJ...468L..91B}.  In the flaring state, the core $\mathrm{T}_{\rm B}$  considerably increases, as also found in previous observations. The highest $\mathrm{T}_{\rm B} > 2.0 \times 10^{11}$ K is detected on 2018 October 4 when the flux density is at the maximum among these datasets. An even higher value $\sim 10^{12}$~K was obtained in \citet{2009ApJ...706..851R} during the 2006 flare. These values approach the equipartition temperature limit for flat-spectrum radio-loud quasars \citep[e.g., ][]{2006ApJ...642L.115H}, suggesting a blazar-like flare in Mrk~231 \citep{2013ApJ...776L..21R}. The peculiar radio core properties (variability, brightness temperature) place Mrk~231 at a position between the radio-loud and radio-quiet AGN populations. 

\subsection{Jet Proper Motion}

\begin{figure}
    \centering
    \includegraphics[width=0.45\textwidth]{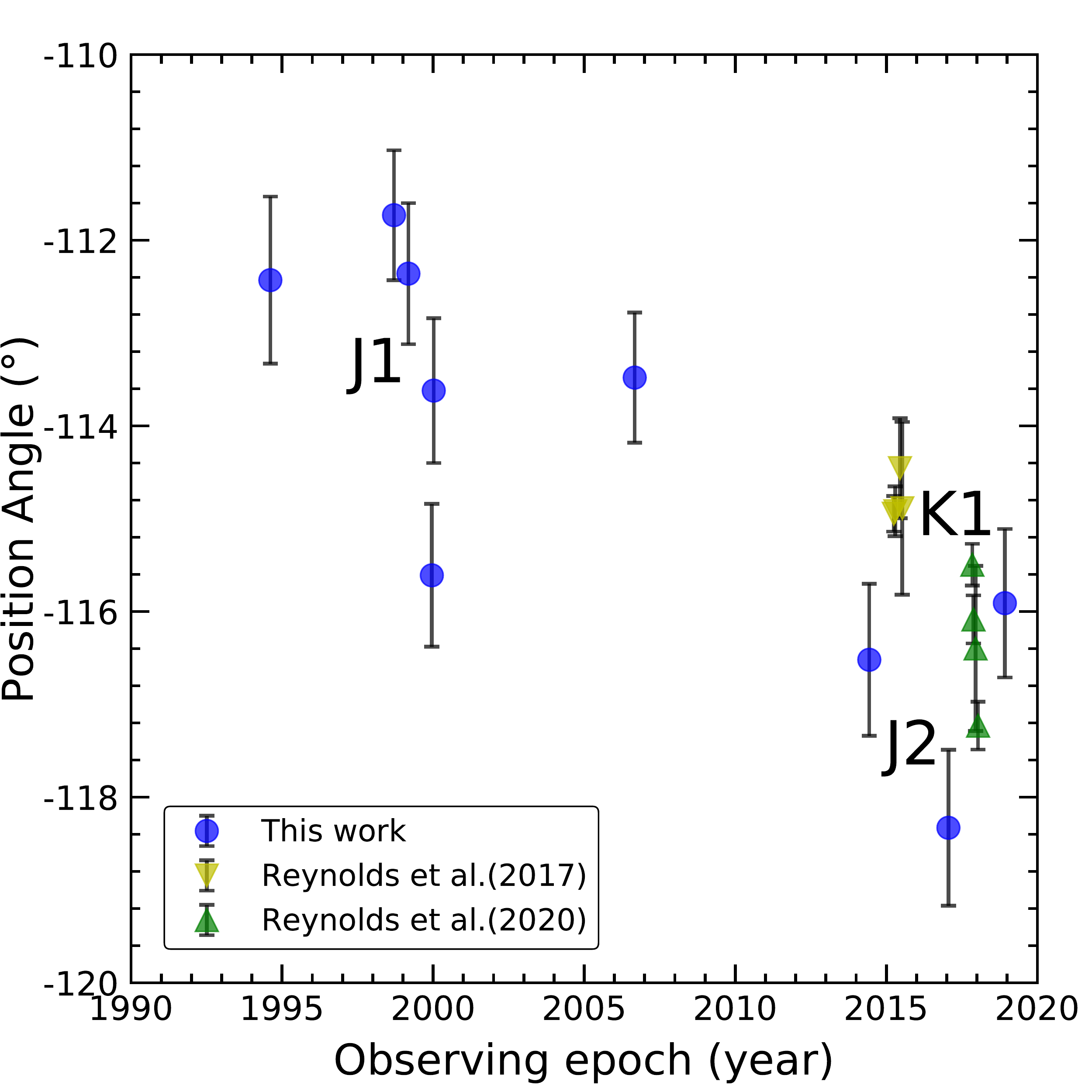}   
    \includegraphics[width=0.45\textwidth]{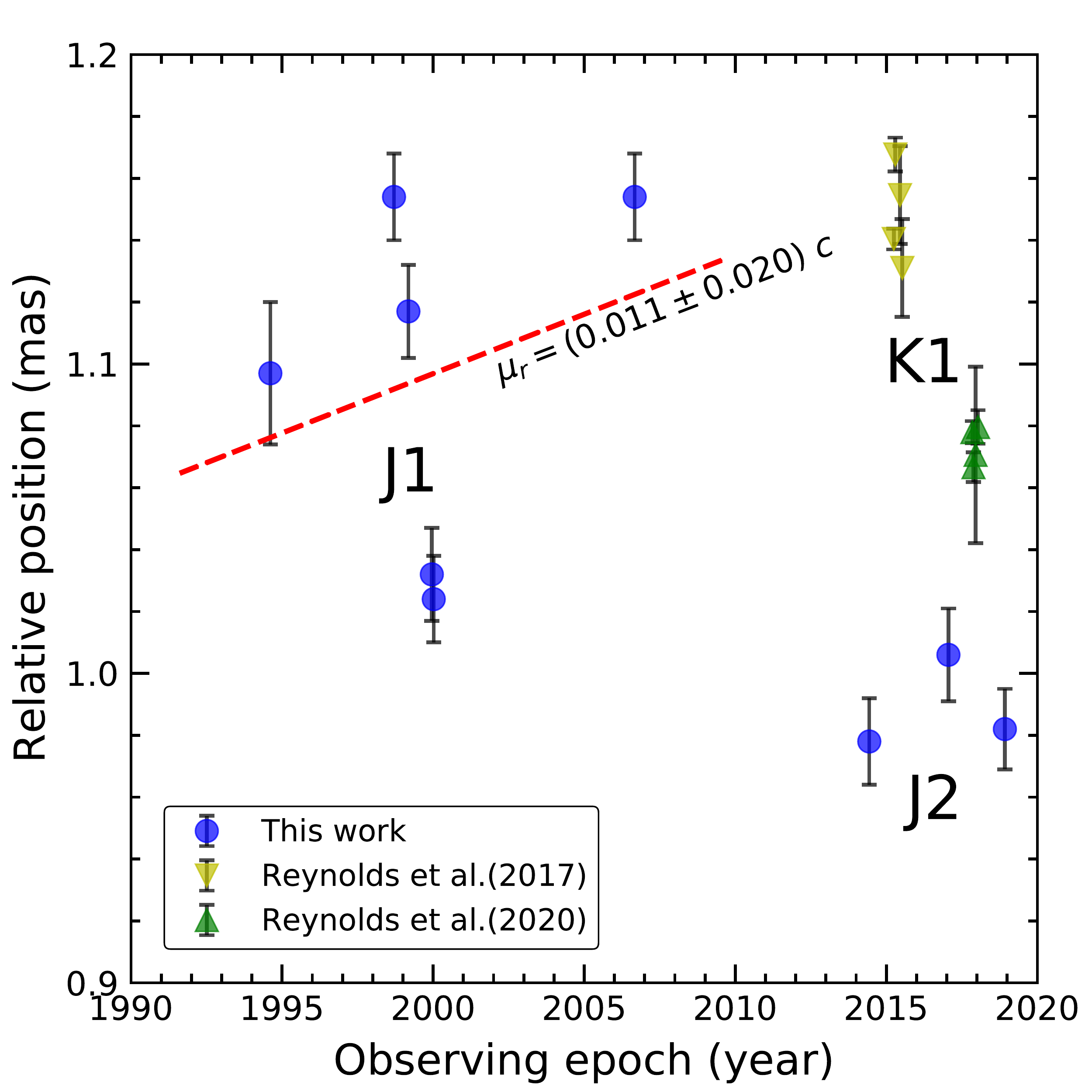}
    \caption{The jet position angle change with time (upper panel) and the relative jet distance with respect to the core versus the observing epoch (bottom panel). Symbols are as follows: blue circles -- Astrogeo database \citep[e.g.,][]{2008AJ....136..580P}, light blue triangles -- \citet{1999ApJ...516..127U}, pink hexagon -- \citet{2017ApJ...836..155R}, and green triangle -- \citet{2020ApJ...891...59R}. }
\label{fig2}
\end{figure}

The bottom panel of Figure \ref{fig2} shows the radial distance of the jet component as a function of time. In addition to the archival Astrogeo VLBI data, we also added the published 8-GHz data points from \citet{2009ApJ...706..851R,2017ApJ...836..155R,2020ApJ...891...59R}. We did not include the higher frequency ($\nu \geq 15$ GHz)  data since the frequency-dependent opacity is likely to induce systematic errors to the core positions. Compared to the previous proper motions made by \citet{2017ApJ...836..155R,2020ApJ...891...59R}, the 8-GHz data used here cover a more extended time baseline of over 14 yr, allowing to determine (or constrain) the kinematics of the underlying jet flow. As we discussed early in Section \ref{sec:3.1}, J1 and J2 have distinctively different position angles, i.e., PA$_{\rm J1} = -113\degr\pm2\degr$, and PA$_{\rm J2} =-117\degr\pm 2\degr$. If they are associated with the same component, the resultant jet proper motion speed would be $-0.005\pm0.002$ mas yr$^{-1}$ ($-0.014\pm0.006\,c$). Negative proper motion lacks a physical ground; therefore J1 and J2 must be different jet components. 

The historical light curve of Mrk~231 shows that the source is in an active state after 2010 \citep{2020ApJ...891...59R}. Several prominent flares are found, peaking in 2011.7, 2013.1, 2015.2 and 2017.9, roughly in a two-year time interval. In contrast, it remained at a relatively lower level of activity from 1994 to 2010. The Mrk~231 jet consists of a slowly moving, continuous background flow and discrete, fast-moving knots which are associated with large episodic flares. The latter might be short-lived and fade rapidly. 

The relative positions of K1 and J2 shows a large scattering (Figure \ref{fig2} bottom panel). The possible reason is that the newly-formed flaring jet component in epochs 2015 and 2017 changed the emission structure in the core region. Since the new jet component is still blended with the core, using the core peak position as the reference point would lead to a systematic error at an uncertain level. For this reason, we do not attempt to calculate the proper motion of J2. 

We use the data points of J1 from 1994 to 2007 to estimate the proper motion. The linear regression fitting gives the proper motion speed, $\mu_{\rm J1} = 0.004 \pm 0.007$ mas yr$^{-1}$ (or $0.011 \pm 0.020\,c$). This small proper motion value is consistent with those of many other radio-quiet AGN jets in magnitude and indicates a subluminal motion. We mention that J1's apparent speed is substantially lower than the previous measurements for the discrete jet knot \citep{2017ApJ...836..155R,2020ApJ...891...59R}.  Our proper motion measurement reinforces the idea that the jet components J1 and J2 detected in the low radio frequency images are associated with the background jet flow. They are distinctive from the superluminal jet knot ejected from the core in the flaring state. 

\section{Discussion}\label{sec:4}

The complex jet structure from the inner pc to outer 25-pc scales indicates morphological and kinematic evolution. The sub-relativistic speed of the background jet flow is suggestive of outflows driven from the inner region, powered by the accretion disc winds \cite[e.g.,][]{2021MNRAS.500.2620Y}. This is especially the case for highly luminous AGN where the bolometric luminosity can be super-Eddington, indicative of a high accretion rate and hence a prominent accretion disc \cite[e.g.,][]{2020MNRAS.494.1744Y,2021MNRAS.500.2620Y}.  The specific scenarios are elaborated below. 

\subsection{Change of the jet structure}\label{sec:4.1}

As shown in Section \ref{sec:3.1}, the position angle of the southern advancing jet has a difference of $60\degr$ from the inner parsec to the outer 25 parsecs. This large change may be enabled by a variety of mechanisms. 
\begin{itemize}
\item 
    deflection of the jet flow caused by a collision between the jet and a massive cloud in the ISM. Such deflections may cause enhanced brightness and polarisation at the sharp bending, especially if the density contrast between the jet and ISM is large  \cite[e.g.][]{2019ApJ...873...11J,2020NatCo..11..143A}. Besides the jet structure change, jet-cloud collisions may also result in deceleration and distortion of the jet structure, as has been seen in the radio-loud quasars 3C 43 \citep[][]{2003A&A...403..537C} and 3C 48 \citep[][]{1991Natur.352..313W,2010MNRAS.402...87A}. Such disruption effects should be even prominent in low-power radio jets (e.g., NGC 1068 \citep{1996MNRAS.278..854M} and NGC 4258 \citep{1991ApJ...381..110P}). However, the southern jet/lobe in Mrk~231 at 25 pc shows a well-defined shape after the presumed collision without any signature of distortion \citep{1999ApJ...516..127U}. This scenario may thus not be operational at these scales.
\item 
    random wandering of the jet head over a large range of position angles. It is possible that the jet impacts on a dense ISM producing an advancing shock which is observed as a radio lobe. The active jet can be deflected by this action and encounter the external medium at different sites until it finds an eventual lower-density location to break through and proceed \cite[e.g.][]{2016A&A...593A..30M}. This wandering behaviour of the jet head is usually used to explain the observed appearance of multiple hotspots in the lobes of the Fanaroff-Riley type II galaxies \citep[e.g., Cygnus A,][]{1985Natur.313...34W}. However, the position angle difference $\sim60\degr$ is too large for the opening angle of a lobe, making this scenario less likely. 
\item 
    rotation of the jet body as influenced by the circum-nuclear disc. This may happen when the jet is not aligned with the rotation axis of the circum-nuclear disc. When the (low-power) jet advances into this disc, a portion of the gas in the disc is entrained into the jet and increases the angular momentum in the perpendicular direction of the jet. The jet body is subject to the differential rotation of the disc through the entrained external gas. This scenario seems possible since the tilted nuclear disc is indeed observed in Mrk~231; the OH megamaser emission \citep{2003Natur.421..821K} reveals a clockwise rotating torus or thick disc located between 30 and 100 pc from the nucleus. The disc axis points at a position angle of $-145\degr$, in a good alignment with the axis of the cone traversed by the curved jet. The observed inner 1-pc and outer 25-pc radio jets coincidentally align with the boundaries between the cone cleaned by the jet and the disc; the mixture and entrainment of external gas into the jet flow at the boundary results in enhanced emission and bending of the jet body following the rotation of the disc. Moreover, the Figure \ref{fig2} upper panel highlights the change of the inner pc-scale jet position angle from $-113^\circ\pm2\degr$ to $-117^\circ\pm2\degr$ in a time interval of $\sim$25 year, providing a hint of the rotation. Continuous monitoring of the VLBI jet structure over a longer time scale would be essential to confirm a smooth and continuous change of the jet direction, as is expected from the rotating jet model. 
\item 
    projection effect of a helical jet. The helical jet flow produces an S-shape in projection on the plane of the sky. The S-shaped structure is seen in nearby Seyfert galaxies \citep[e.g., NGC 4258, ][]{2005ApJ...622..178G} and in small-sized GHz-peaked spectrum (GPS) galaxies \citep[e.g.,  B0500+019,][]{2001A&A...377..377S}. The driving mechanism of helical jets may include a regular precession of the jet axis  \cite[e.g.,][]{2005A&A...431..831L,2012MNRAS.421.1861V} near the base enabled by a sub--pc-scale SMBH binary system \cite[e.g.,][]{1980Natur.287..307B,2016MNRAS.463.1812M}, a tilted or warped disc caused by a spinning black hole \citep[e.g., ][]{1975ApJ...195L..65B,1992MNRAS.258..811P,2005MNRAS.363...49K,2005ApJ...635L..17L},  magneto-hydrodynamic instabilities \cite[e.g.,][]{1987ApJ...318...78H}, or due to developing disc-jet instabilities \cite[e.g.,][]{2003ApJ...584..135L,2010MNRAS.402...87A,2013MNRAS.434.3487A}. For a simple ballistic motion of the precessing jet, the twisted trajectory is a projection of the jet knots ejected at various position angles \citep{1981ApJ...250..464L}. It is likely that the component J1 (1994 -- 2007), the component J2 (2014 -- 2019), and the 25-pc southern lobe, which are at different position angles, are indicative of such episodic behaviour. A SMBH binary model was proposed by \citet{2015ApJ...809..117Y} to account for the deficit of near-UV continuum flux in Mrk 231. If the helical jet is driven by the tidal interaction between the primary and secondary black holes, the inferred precession period is on the order of $10^3$ yr and the mass ratio is less than 0.03, \textit{i.e.}, an intermediate-mass black hole orbits the primary $10^8\, M_\odot$  one. However, the presence of binary black holes lacks convincing observational evidence \citep[e.g.,][]{2016ApJ...825...42V,2016ApJ...829....4L}. Owing to the data used (PA variation) being sparse and spanning a very short time span, a precession of the jet can not be confirmed from the currently available VLBI data.  . 
\end{itemize}

In summary, the first two models (\textit{i.e.}, deflection due to jet-cloud collision, and random wandering of jet head) lack strong observational support from the existing data. High sensitivity VLBI polarimetric observations of Mrk~231 are necessary to explore the location of the jet bending and the polarisation structure to confirm or exclude these scenarios. Other mechanisms (jet rotation and helical jet) seem possible but warrant further  observations and detailed studies. In either scenario, the interactions between the bending jet and the ISM result in large-aperture-angle shocks that sweep the ISM within a cone and expel the gas to move outward. This dynamic process happens at the onset of the large-scale ionised and molecular outflows and may participate in powering the outflows (see more discussion below).

\subsection{Stagnation of the low-power jet in the host galaxy}\label{sec:4.2}

Mrk~231 lacks a large-scale radio jet structure beyond a few kpc \citep{1998AJ....115..928C,1999ApJ...516..127U,1999ApJ...519..185T,2016A&A...593A..30M}. Although there is some extended emission a few hundred mas south of the nucleus, those diffuse features might not be directly related to the jet \citep{2016A&A...593A..30M}. The compact triple structure has a projected extent of $\sim 50$ mas ($\sim 40$ pc) in the north-south direction. The southern lobe is brighter and longer (25 pc in projection), suggesting it is associated with the advancing jet. The sub-kpc source size and symmetry morphology is reminiscent of a compact symmetric object (CSO). The radio power of Mrk~231 is $\log P_{\rm 1.4GHz} = 24.15$ W Hz$^{-1}$ \citep{2016A&A...593A..30M}, placing it in the low-power CSO population \citep{2012ApJ...760...77A}. 
The subluminal jet speed measured for the background jet in Mrk~231 is consistent with its low-power nature. The proper motion vector component of the jet head in the south direction is $0.011\, c$. If we take this value as the upper limit of the lobe expansion speed, we can estimate the age of the radio structure to be $\gtrsim$7400 yr. The radio structure's actual age could be much longer than this value due to the obstruction of the jet growth. 

The jet structure strongly influences the size and power of the radio source. When the jet head wiggles over a large range of directions, it continuously hits new regions of the ISM. At each direction, the time-averaged jet power is small. 

\begin{figure}
    \centering
    \includegraphics[scale=0.35]{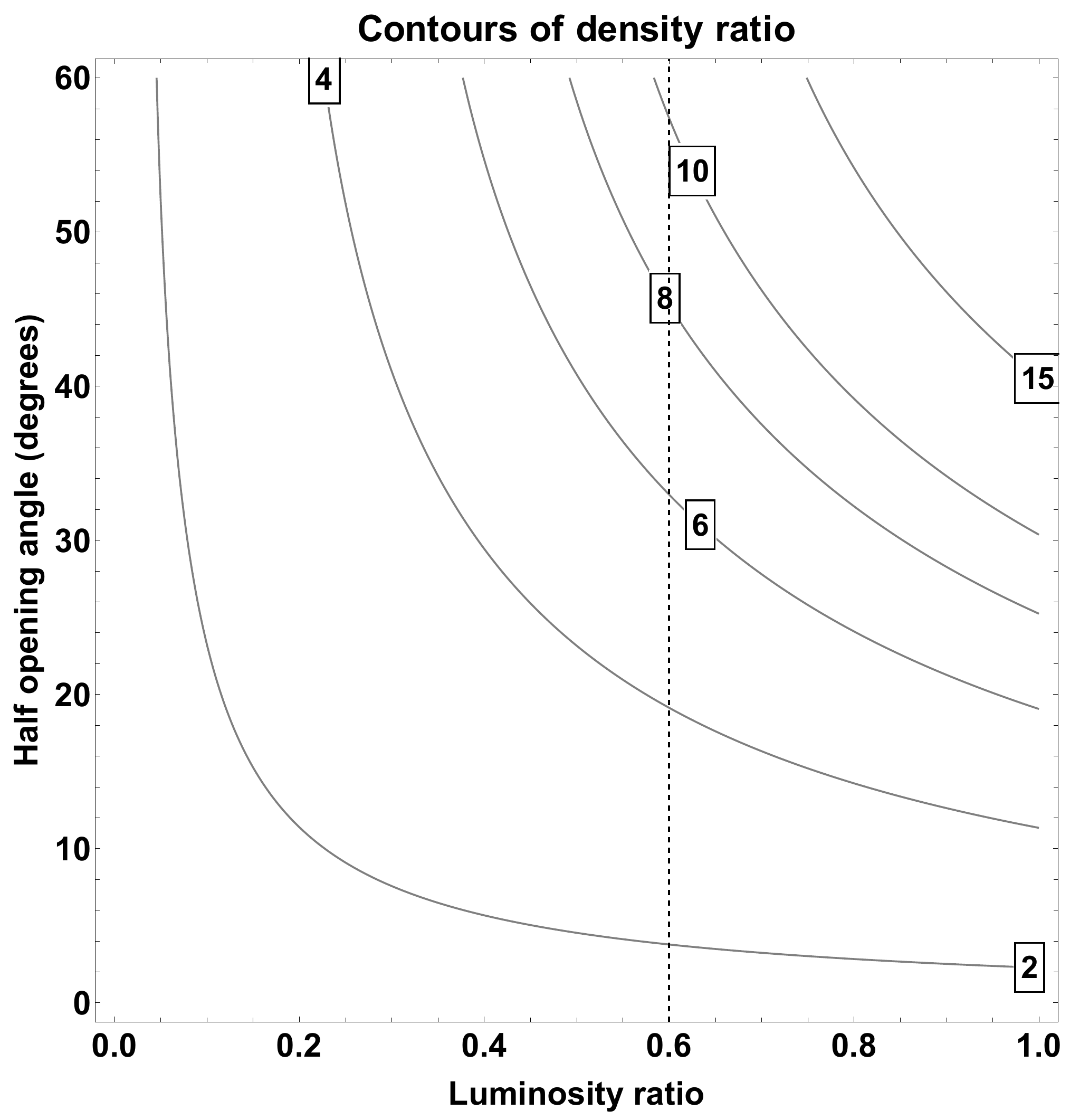}
    \caption{Contours of the density contrast as a function of the luminosity ratio $L_{\rm j,kin}/L_{\rm j}$ and jet half opening angle $\theta_0$. The vertical dashed line marks $L_{\rm j,kin}/L_{\rm j} = 0.6$ for Mrk 231 (see discussion in \ref{sec:4.3}) indicating $\kappa \gtrsim 1.2$.}
    \label{kappathetaL1}
\end{figure}

The jet power ($L_{\rm j}$) is composed of radiative power ($L_{\rm j,rad}$) and kinetic power ($L_{\rm j,kin}$). The study of \citep{2014IJMPS..2860188F} employs a statistically viable sample of jetted AGN for a resulting empirical relationship between the constituent jet luminosities $L_{\rm j,rad}$ and $L_{\rm j,kin}$ and the core radio luminosity $L_{\rm radio}$,
\begin{align}
    \log L_{\rm j,rad} &= (12\pm 2) + (0.75\pm 0.04)~\log L_{\rm radio}\\ \nonumber
    \log L_{\rm j,kin} &= (6\pm 2) + (0.90\pm 0.04)~\log L_{\rm radio}.
\end{align}
Owing to the small physical separation between the jet and core components, and the unresolved nature of the latter, the putative radio core may be brighter. We then take the combined flux density of $\approx$ 400 mJy during the 2018 epoch, an upper limit as representative of that from the radio core. The corresponding radio luminosity is about $1.5\times10^{41}$erg s$^{-1}$, which converts to $\log L_{\rm j,rad} = 42.89$ and $\log L_{\rm j,kin} = 43.06$ using the above equations and hence, a luminosity ratio $L_{\rm j,kin}/L_{\rm j} = L_{\rm j,kin}/(L_{\rm j,kin}+L_{\rm j,rad})$ of 0.6. The balance of momentum flux between the jet and external ISM on interaction and the density contrast can be cast in terms of the kinematic, geometric, and energetic parameters governing the jet and ISM \cite[e.g.,][]{1998ARA&A..36..539F}. 
\begin{align}\label{mombal1}
    &\frac{L_{\rm j}}{(\beta_j-\beta_h) c} = \kappa \rho_j \beta^2_h c^2 A_h   \\ 
    &\kappa = \left(\frac{\beta_j}{\beta_h}-1\right)^2, \label{mombal2}
\end{align}
where $\beta_j$ is the speed of the jet flow, $\beta_h$ is the speed of the advancing jet head at the working surface, and $A_h = r^2_h \sin \theta_0$ is the surface area in contact at the working surface of size $r_h$ and for a jet half opening angle $\theta_0$, and $\kappa$ is the density contrast between the external ISM $\rho_{\rm ISM}$ and the jet flow $\rho_j$ \citep{2020NatCo..11..143A}. The density of the baryonic jet is
\begin{equation}
    \rho_j = \frac{L_{\rm j, kin}}{(\pi r^2/2) \beta^3_j c^3 \Gamma^2_j},
\label{rhojet}
\end{equation}    
where $r$ is the jet extent and Lorentz factor $\Gamma_j = (1-\beta^2_j)^{-1/2}$. The equations (\ref{mombal1}), (\ref{mombal2}) and (\ref{rhojet}) are used along with $\sin \theta_0 \approx r_h/r$ to cast the density contrast as
\begin{equation}
    \kappa = \left(1-\frac{2}{\pi} \left(\frac{L_{\rm j, kin}}{L_{\rm j}}\right)^{1/3} \frac{\sin \theta_0}{\Gamma^{2/3}_j}\right)^{-2},
\end{equation}
which after some algebraic manipulation can be numerically solved for the density contrast $\kappa$ as
\begin{equation}
    C \beta^2_h x^5+2 C \beta^2_h x^4+(1-C+C \beta^2_h) x^3+3 x^2+3 x+1 = 0,    
\end{equation}
where $C = \left(\frac{2}{\pi} \frac{L_{\rm j,kin}}{L_{\rm j}}\right) \sin^3 \theta_0$ and $x = \kappa^{1/2}$. The contrast ratio $\kappa$ is now a function of jet properties that can be directly inferred from the VLBI observations or constrained. The measured $\beta_h = 0.013 \ c$ is used in the above equation to solve for $\kappa = \kappa \left(L_{\rm j,kin}/L_{\rm j},\theta_0\right)$ and the contours are plotted in \autoref{kappathetaL1}. The luminosity ratio is taken in the range $0.01 - 1.0$ and the jet half opening angle in the range $0.1^\circ - 60^\circ$ for the numerical solution. We obtain $\kappa$ in the range $1.1 - 28.6$ for the above range, indicating a generally denser surrounding medium. This argument of dense external ISM is consistent with the OH megamaser observations \citep{2003Natur.421..821K}. $\kappa$ is larger for wider opening angles, consistent with the expectation for a weaker, relatively un-beamed jet/outflow. 

Comparing the jet radiative power and kinetic power shows that a substantial fraction of the jet power is in the form of radiation. The jet-ISM interactions lead to a bulk of jet mechanical energy dissipation to the shocks, which in turn power and expel the large-scale nuclear outflow winds/outflows. As a result, the advancing motion of the jet flow is likely to inhibit or slow down. Thus, in a rotating, clumpy, dense ISM environment such as in Mrk~231, the growth of the low-power radio source could be stopped at some distance, as predicted by the frustrated model for compact radio sources \citep{1984AJ.....89....5V,1997ApJ...485..112B,2000MNRAS.319..445S}.

\subsection{Jet contributes to powering the wide-angle outflows}\label{sec:4.3}

Wide-angle outflows on sub-kpc to kpc scales are ubiquitous in radio-quiet quasars \citep{2005ARA&A..43..769V}. Sustenance of the wide-angled outflows at large scales requires the AGN energy and momentum to be transferred into the ISM over a substantial volume through radiative or mechanical processes \citep{2011ApJ...729L..27R,2012ApJ...750...55S}. Radio jets can enable this feedback and facilitate the expulsion of neutral and ionised gas from the nuclear region \citep{2005A&A...444L...9M,2009MNRAS.400..589H,2009ApJ...690..953F,2020arXiv200102675M}.

Mrk~231 hosts powerful multi-phase outflows extending at multiple scales, including the nuclear ultra-fast X-ray winds, ionised gas outflows arising from the broad-line region, and the galactic-scale molecular and atomic gas outflows \citep{2015A&A...583A..99F,2016A&A...593A..30M}. The inferred jet flow speed in this work is slightly lower than the X-ray ultra-fast winds (up to 20000 km s$^{-1}$). The inverse correlation between the radio and far-UV luminosity in Mrk~231 indicates a direct connection between the jet and the BAL winds \citep{2017ApJ...836..155R}. 

Unlike the radio-loud quasars which usually have collimated and highly relativistic jet, the jets in radio-quiet quasars are less powerful and uncollimated. The subrelativistic jets are often regarded as winds or outflows \citep{1992ApJ...396..487S}.  Large-scale (up to $\sim$10 kpc) loosely collimated outflows are found to be ubiquitous in radio-quiet Type 1 quasars, mostly along the minor axes of the host galaxies, and are thought to be quasar-driven \citep[e.g., ][]{2017ApJ...850...40R}. Mrk~231 sits between the radio-loud and radio-quiet quasars. It is one of the few radio-quiet Type 1 quasars having prominent jets; another such example is PG 1700+518 which shows strong evidence of a jet-driven outflow on kpc scales \citep{2012MNRAS.419L..74Y,2018ApJ...852....8R}. The PG 1700+518 jet lies along with the outflows. Similarly, the Mrk~231 jet and the BAL ionised outflows exist on the same scale. The change of the pc-scale jet structure over a broad range of directions results in advancing shocks that can participate in ionising and accelerating the BAL winds, and in the expulsion of gas at larger scales. The time-averaged kinetic energy flux of the jet is $\sim 10^{41-42}$ erg s$^{-1}$ in the quiescent state, and increases to $\sim 10^{43}$ erg s$^{-1}$ during the flaring state \citep{2009ApJ...706..851R}, which accounts for (1--35)\% of the energy rate of the molecular outflow \citep{2015A&A...583A..99F}. These clues suggest that the contribution of the jet to powering the outflows is not negligible when Mrk~231 is in its flaring state.

\section{Summary}

We analysed archive VLBI data of Mrk 231 and presented the milli-arcsecond resolution images that reveal the inner parsec jet structure. Comparing the published data with ours, we found a large misalignment ($\sim 60\degr$) between the inner pc jet and the outer 25-pc jet. A hint of jet position angle change is seen in the inner pc scale and warrants new observations to confirm. The jet structure can be explained by a scenario in which the jet lies in a large opening angle cone and expands in various directions. The cone swept by the jet is likely to be physically associated with the dusty disc revealed from the previous OH megamaser observations. The jet knots revealed in the 8-GHz VLBI images represent the active working surface of the jet head hitting on the external ISM. The entrainment of the gas from the rotating circum-nuclear disc in the jet may result in the rotation of the jet body, or the jet itself follows a helical trajectory driven by jet precession or other mechanisms. The expansion of the jet within the cone creates large-aperture-angle shocks that accelerate and expel material from the nuclear region. The propagation of the jet energy and momentum outwards further powers the molecular and atomic gas outflows. Although the quasar radiation might be the dominant power source of the AGN winds/outflows, the jet power accounts for a  fraction of the energy rate of the molecular outflow in Mrk 231. Thus, the Mrk 231 jet's contribution to the feedback to its nuclear environment, especially in the flaring state, may not be negligible. The jet direction change within the large-opening-angle cone causes the time-averaged kinetic energy and momentum in a specific direction to be significantly lower than a collimated jet flow. Moreover, a substantial fraction of the jet power is dissipated in radiation. Consequently, the Mrk~231 jet is obstructed by the dense ISM within a few tens pc scale and fails to grow into a large-sized radio source.  

\section*{Acknowledgements}

We thank the anonymous referee for her/his constructive comments which greatly improved the manuscript. This work is supported by the National Key R\&D Programme of China (2018YFA0404603) and the Chinese Academy of Sciences (CAS, 114231KYSB20170003). S.J. is supported by the CAS-PIFI (grant No. 2020PM0057) postdoctoral fellowship. Y.C.W. thanks the hospitality of Shanghai Astronomical Observatory during her summer internship. The authors acknowledge the use of Astrogeo Center database maintained by L. Petrov. The National Radio Astronomy Observatory are facilities of the National Science Foundation operated under cooperative agreement by Associated Universities, Inc. 

\section*{Data availability}

The datasets underlying this article were derived from the public domain in NRAO archive (project codes: BU013, BA080;
\url{https://science.nrao.edu/observing/data-archive}) and Astrogeo archive (project codes: BB023, RDV13, BG219D, UF001B, UG002U;
\url{http://astrogeo.org/}). The calibrated visibility data can be shared on reasonable request to the corresponding author.




\bibliographystyle{mnras}
\bibliography{Mrk231} 





\bsp	
\label{lastpage}
\end{document}